# Fast magneto-ionic switching of interface anisotropy using yttria-stabilized zirconia gate oxide

Ki-Young Lee[1†], Sujin Jo[1†], Aik Jun Tan[2], Mantao Huang[2], Dongwon Choi[1], Jung Hoon Park[3], Ho-Il Ji[3,4], Ji-Won Son[3,5], Joonyeon Chang[1,6], Geoffrey S. D. Beach[2*] and Seonghoon Woo[1*††]

[1]Center for Spintronics, Korea Institute of Science and Technology, Seoul 02792, Korea

[2]Department of Materials Science and Engineering, Massachusetts Institute of Technology, Cambridge, MA, USA

[3]Center for Energy Materials Research, Korea Institute of Science and Technology, Seoul 02792, Korea

[4]Nanomaterials Science and Engineering, Korea University of Science and Technology (UST) KIST Campus, Seoul 02792, Korea

[5]Graduate School of Energy and Environment (KU-KIST GREEN SCHOOL), Korea University, Seoul 02841, Korea

[6]Department of Materials Science & Engineering, Yonsei University, Seoul 03722, Korea

[†] These authors contributed equally to this work.

[††] Present address: IBM T.J. Watson Research Center, 1101 Kitchawan Rd, Yorktown Heights, New York 10598, USA

* Author to whom correspondence should be addressed: gbeach@mit.edu, shwoo@ibm.com

**Abstract**

Voltage control of interfacial magnetism has been greatly highlighted in spintronics research for many years, as it might enable ultra-low power technologies. Among few suggested approaches, magneto-ionic control of magnetism has demonstrated large modulation of magnetic anisotropy. Moreover, the recent demonstration of magneto-ionic devices using hydrogen ions presented relatively fast magnetization toggle switching, $t_{sw}$ ~ 100 ms, at room temperature. However, the operation speed may need to be significantly improved to be used for modern electronic devices. Here, we demonstrate that the speed of proton-induced magnetization toggle switching largely depends on proton-conducting oxides. We achieve ~1 ms reliable (> $10^3$ cycles) switching using yttria-stabilized zirconia (YSZ), which is ~ 100 times faster than the state-of-the-art magneto-ionic devices reported to date at room temperature. Our results suggest further engineering of the proton-conducting materials could bring substantial improvement that may enable new low-power computing scheme based on magneto-ionics.

Although non-volatile spintronics-based devices are seen as candidates to replace conventional CMOS technologies[1], electrical current-induced magnetization switching is energy-intensive[2], which has hindered the implementation of spintronics in low-power applications. For example, although recent developments in spin-transfer-torque magnetoresistive random access memory (STT-MRAM) have reduced the switching energy to ~ 100 fJ/bit[3,4], the energy consumption is still much higher than that of conventional static-RAM (SRAM), which STT-MRAM is being targeted to replace in some applications[2]. Voltage-control of magnetic anisotropy[5,6], in which a gate voltage applied across a thin oxide layer modulates the magnetization switching energy barrier in an adjacent ferromagnet, has attracted a significant attention in recent years, owing to the much reduced switching energies in devices. Although multiple approaches have been demonstrated experimentally[2,7–16], voltage-controlled ionic motion leading to magnetic anisotropy modulation, known as magneto-ionic control of magnetic anisotropy[9–16], has yielded high-efficiency anisotropy modulations up to ~1 erg cm$^{-2}$ [10,11], as required for practical device applications.

Early works[9–14] utilized voltage-induced oxygen ion ($O^{2-}$) migration in ferromagnetic heterostructures, however, such a scheme relying on reactive and relatively large oxygen ions suffered from slow switching speed[10,17], irreversible oxidation damage[13] and high operation temperature[10,11]. For example, the control of interfacial magnetic anisotropy using $O^{2-}$ ions requires >100 s at room temperature, thereby requiring high temperature of > 250 ºC in order to reduce the switching speed down to ~ 30 s[17]. Recently, it has been demonstrated that proton ($H^+$)-based magneto-ionic devices could offer significantly enhanced device capabilities: non-destructive ~100 ms-speed magnetization switching at room temperature (RT), with robust, reversible operation (proton injection and release) for thousands of switching cycles[15,16]. In these devices, solid-state electrochemical water splitting using atmospheric humidity was shown to provide a source of mobile protons that could be driven through a gate oxide to reversibly switch the magnetic anisotropy between perpendicular magnetic anisotropy (PMA) and in-plane magnetic anisotropy (IMA)[15,16]. Considering the nanosecond timescale required for modern data storage and computing devices, however, significant improvements still need to be made. One possible approach is to use an optimized proton-conducting oxide in order to improve the speed of the device, since previous studies have only focused on a single material, GdOx[15,16], which is not generally considered a good proton conductor.

Here, we show that the speed of proton-induced magnetization switching depends largely on the proton conductivity of the proton-conducting oxide material. By comparing various complex oxides known for their efficient proton conductivity[18], namely gadolinium-doped ceria (GDC), barium cerium yttrium zirconate (BZCY) and yttria-stabilized zirconia (YSZ), we demonstrate fast (~1 ms) and reliable (> $10^3$ cycles) magnetization switching in ultrathin ferromagnetic heterostructures employing YSZ gate oxides. The observed speeds are two orders of magnitude faster than the state-of-the-art device reported to date at RT. Systematic material and thickness studies reveal a path toward optimization of the device structure for efficient electrochemical interface reactions and fast ion transport, which may push the switching speed toward levels suitable for data handling device applications.

Proton conducting oxides have been widely studied in the context of solid oxide fuel cells (SOFCs), where high proton mobility at low-temperature (T < 500 ºC) is required for efficient device implementations[18–21]. Among various classes of oxides, enhanced ionic conductivity has been achieved in fluorite-structure doped $ZrO_2$ and $CeO_2$, such as yttria-stabilized $ZrO_2$[22] and gadolinium-doped $CeO_2$[23], as well as perovskite-structured mixed oxides such as (yittria-doped) $BaCeO_3$-$BaZrO_3$[24]. These oxides are stable under electrochemical reactions (i.e. reduction or oxidation) and present large (small) ionic (electronic) conductivities. These same characteristics make them particularly suitable as proton conductors in proton-based magneto-ionic devices. In this study, we examine magneto-ionic switching in heterostructures employing such protonic oxides as gate electrodes.

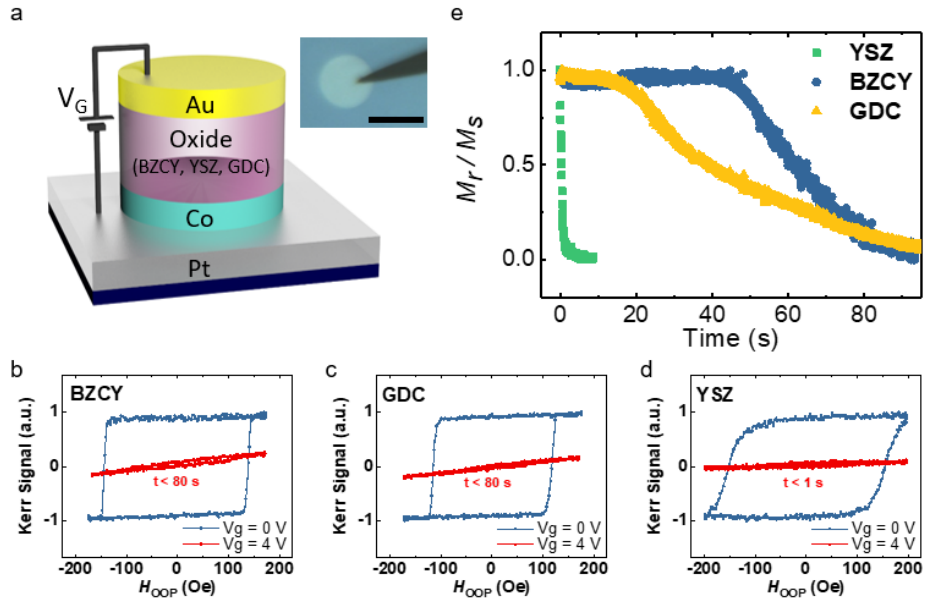

**Figure 1. Magneto-ionic switching of ferromagnetic heterostructures with various oxides.**
**(a)** Schematic of magnetic heterostructure of Pt(3 nm)/Co(1 nm)/Oxides(10 nm)/Au (3nm) for MOKE measurements. The inset shows an optical micrograph of the probe-landed device. Scale bar, 50 μm. Out of plane hysteresis loops measured in the devices with **(b)** BZCY, **(c)** GDC and **(d)** YSZ, respectively, under gate voltage $V_G = 0$ V (blue line) and $V_G = +4$ V (red line). **(e)** Out-of-plane magnetization remanence ratio, $M_r/M_s$, measured by MOKE, where $M_r$ is the remnant magnetization and $M_s$ is the saturation magnetization as a function of gate voltage application time at $V_G = +4$ V, corresponding to BZCY (blue), GDC (yellow) and YSZ (green) gate oxides.

Figure 1 summarizes the effect of gate voltage application to Pt(3 nm)/Co(1 nm)/Oxide(20 nm)/Au(10 nm) ferromagnetic heterostructures with perpendicular magnetic anisotropy, where GDC, BZCY and YSZ were used as gate oxides [see Methods for device fabrication details]. Figure 1a shows the schematic of the sample structure and an optical micrograph of a 50 μm-diameter device. A gate voltage $V_G$ = +4 V was applied to the top Au electrode using a BeCu microprobe while the bottom Pt was grounded. The hysteresis behavior of magnetic heterostructures was characterized by polar magnetic-optic Kerr effect (MOKE) measurements, as shown in Figs. 1b-d. Note that all magnetic heterostructures exhibit PMA with similar coercive fields in the as-grown states, implying that PMA may originate from strong interfacial spin-orbit coupling between the Pt and Co layers. Moreover, prolonged application of positive gate-bias eventually eliminated the PMA in all devices. This can be attributed to proton-induced magnetization rotation to an in-plane state as summarized in Fig. 2. This result is consistent with recent observation using GdOx[15,16], revealing the generality of this mechanism. However, the switching time strongly depends on the type of oxide. Figure 1e plots the variation of out-of-plane remnant magnetization ratio ($M_r/M_s$) as a function of gate-bias application time. Surprisingly, when YSZ was used as the gate oxide, $M_r/M_s$ approaches zero within a timescale $t_{sw}$ < 1 s, which is approximately two orders of magnitude faster than for BZCY and GDC. This result implies fast ionic transport in 20 nm-thick YSZ gate oxide grown at room temperature, which likely originates from enhanced proton conductivity through e.g. grain boundaries[22,25], which is also largely material and growth condition dependent[26].

To verify the underlying mechanism as proton-induced magnetic anisotropy modulation, we performed simultaneous polar- and longitudinal-MOKE and cyclic voltammetry (CV) measurement as shown in Fig. 2. Figure 2a shows the schematic description of proton-induced magnetic anisotropy switching. In this scheme, as was briefly explained in the introduction, upon the application of positive voltage exceeding ~1.2 V, atmospheric water (humidity) is electrolyzed at the top electrode to produce $H^+$ and $O_2$ gas. The generated protons are transported to bottom electrode through the oxide via a Grotthuss-type hopping mechanism[15,20], where their presence modulates interfacial magnetic anisotropy at YSZ/Co and Co/Pt interfaces. Figures 2b-e present out-of-plane and in-plane hysteresis loops for Pt/Co/YSZ heterostructures, measured by polar and longitudinal MOKE. The MOKE geometry was switched between polar and longitudinal configurations (Figs. 2b-c) to successively probe the out-of-plane and in-plane hysteresis behaviour every 10 s. In the virgin state (black line in Figs. 2d, e), square and linear loops are measured using polar and longitudinal MOKE respectively, implying an initial PMA state.

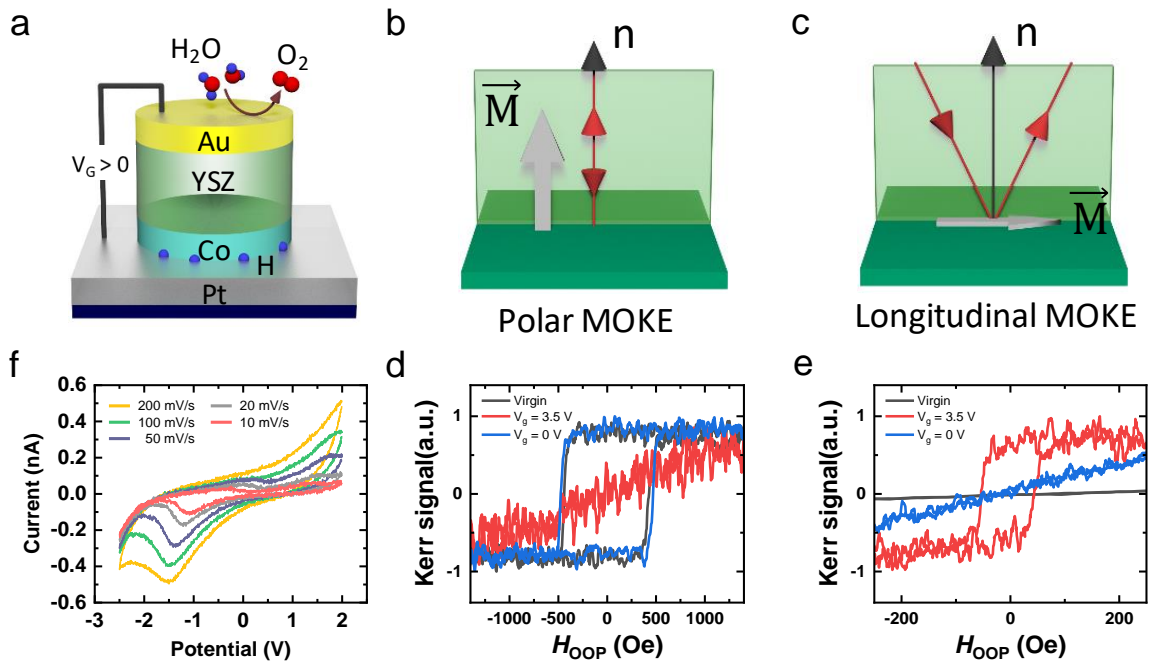

**Figure 2. Proton-induced magnetization rotation in the structure using YSZ gate oxide.** **(a)** Schematic of magneto-ionic switching mechanism in a Pt(3 nm)/Co(1 nm)/YSZ(10 nm)/Au (3nm) structure device. **(b)-(c)** Schematics of polar and longitudinal MOKE measurement configurations, respectively. **(d)-(e)** Polar MOKE and longitudinal MOKE hysteresis loops in the virgin state (black line), after applying $V_G = +3.5$ V (red line), and after $V_G$ is removed (blue line). **(f)** Cyclic voltammetry measurement plot at various sweep rates performed in an ambient atmosphere using a Au(10 nm)/YSZ(50 nm)/Pt(3 nm) structure.

However, upon application of $V_G = +3.5$ V, the magnetization rotates in plane (red line in Figs. 2d, e), corresponding to hydrogen accumulation at the YSZ/Co and Co/Pt interfaces. When the gate voltage set to zero (short-circuiting the top and bottom electrode), PMA spontaneously recovers and remains stable (blue line in Figs. 2d, e), as accumulated protons at the interfaces are removed[15]. We then performed CV measurement for a Au(10 nm)/YSZ(50 nm)/Pt(3 nm) structure at different sweep rates, as shown in Fig. 2f. Although cathodic and anodic currents were observed for all sweep rates, the magnitude of the current was governed by the voltage sweep rate, suggesting that the process is mass transport-limited rather than charge-transport limited[27]. Therefore, the polarity of gate voltage used for magnetization rotation and the mass transport nature revealed from CV data are consistent with voltage-induced proton pumping inferred from the magnetic measurements heterostructures, as also reported elsewhere[15].

Having examined the mechanism and observed that proton-induced magnetization switching can take place in YSZ-based magnetic heterostructures efficiently, we then demonstrate the fast and robust switching of magnetization in further engineered YSZ-based heterostructures. Figure 3a shows the schematic of a Pd(3 nm)/Co(1 nm)/Pd(1 nm)/YSZ($t_{YSZ}$ nm)/Pt(3 nm) structure used for fast switching experiments, where $t_{YSZ}$ is controlled between 3 nm and 10 nm. Compared to the previous structures used for Fig. 1 and Fig. 2, we first substituted Au with Pt for the top electrode, as Pt is known as an excellent catalyst for oxygen evolution reaction (OER)[28]. This supposition is supported by our direct comparison between Au and Pt on cyclic voltage application experiments, as shown in Supplementary Fig. S1. Moreover, we introduced a thin layer of Pd as a hydrogen loading layer[29] between Co and YSZ, where the addition of Pd is expected to protect Co from degradation over many switching cycles [see Supplementary Fig. S2 and Fig. S3]. We also replaced the bottom Pt reference electrode with a Pd layer, so that a larger quantity of hydrogen can be absorbed at the Pd/Co interface, inducing more efficient change in interfacial magnetic anisotropy from the Pd/Co interface. Figure 3b plots the change of out-of-plane remnant magnetization ratio ($M_r/M_s$) upon $V_G$ = +6 V gate voltage application on a $t_{YSZ}$ = 6 nm sample, and Figs. 3c-d plot the measured out-of-plane hysteresis loops right before (Fig. 3c) and right after (Fig. 3d) the gate voltage application. In this measurement scheme, we sweep the out-of-plane magnetic field and measure $M_r/M_s$ via hysteresis loops measured at a field sweep rate of 500 Hz, corresponding to a time resolution of 2 ms. As shown in Figs. 3b-d, upon gate voltage application, the magnetization rotates from PMA to in-plane instantaneously within the time resolution of current setup. This fast switching in the millisecond regime is two orders of magnitudes faster than the state-of-the-art magneto-ionic device reported to date at room temperature, which was recently demonstrated with a $GdO_x$ gate oxide using proton-induced magneto-ionic switching[15].

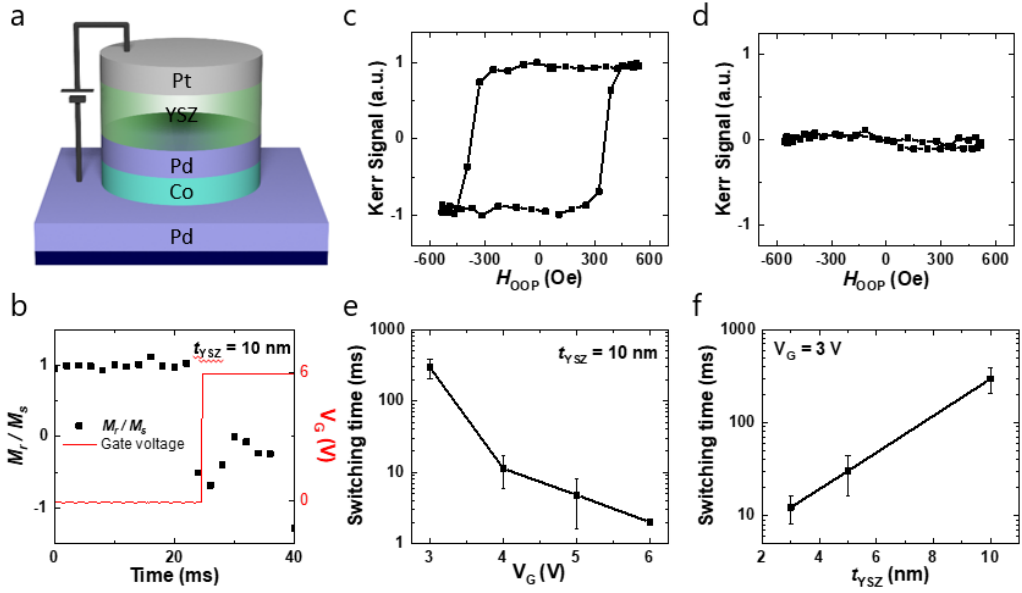

**Figure 3**. **Fast switching of magnetic anisotropy in an engineered structure with YSZ gate oxide.** **(a)** Schematic of device with optimized structure for fast magnetization switching, Pd(3 nm)/Co(1 nm)/Pd(1 nm)/YSZ($t_{YSZ}$ nm)/ Pt(3 nm), where $t_{YSZ}$ is controlled between 3 nm and 10 nm. **(b)** Polar MOKE measurement of the remanence ratio ($M_r/M_s$) upon gate voltage application, $V_G$ = +6 V. Out-of-plane hysteresis loops right **(c)** before (where $V_G$ = 0 V) and **(d)** after (where $V_G$ = +6 V) the gate voltage application. Magnetization switching time, $t_{sw}$, as a function of **(e)** gate voltage amplitude, $V_G$, and **(f)** YSZ gate oxide thickness, $t_{YSZ}$, respectively.

Figure 3e plots the effective magnetization switching time, $t_{sw}$, as a function of gate voltage amplitude at $t_{YSZ}$ = 10 nm, where $t_{sw}$ exponentially decreases from $t_{sw}$ = 290 ms ($V_G$ = +3 V) to $t_{sw}$ ~ sub-2 ms ($V_G$ = +6 V) as the gate voltage increases. This exponential decrease suggests a mobility-limited switching time, which is expected to scale exponentially with the electric field[30]. The electric field can also be varied by controlling the YSZ thickness. As seen in Fig. 3f, the switching time varies exponentially with gate oxide thickness, for a fixed gate voltage $V_G$ = +3 V, corresponding to a value appropriate for applications. It should be noted that our experiments use a mechanical probe to contact the top electrode, which limits the YSZ thickness to > 3 nm to prevent mechanical shorting; such limitations can be eliminated using microfabricated top gate contacts, such that even thinner gate oxides and consequently faster operation can reasonably be expected. From Fig. 3f, we estimate that an effective YSZ thickness below 1 nm could bring the operation speed down to µs and even ns regime in the current scheme.

We next examine the endurance of this YSZ-based magneto-ionic device with $t_{YSZ}$ = 10 nm device, as summarized in Fig. 4. Figures 4a, b present $M_r/M_s$ extracted from polar MOKE hysteresis loop measurements as a function of voltage cycle, for cycles 1–10 and 1001–1010, tracking in-plane/out-of-plane transitions as $V_G$ is cycled between +4 V and −1 V. The asymmetric voltage amplitudes are chosen to maintain the high speed of operation at both loading/unloading processes, while minimizing oxidation of Co, which occurs with repeated cycling at higher negative bias. Figures 4c, d show corresponding magnetization hysteresis loops measured by polar MOKE at cycle 10 and cycle 1010, respectively, revealing the robust reliability of YSZ-based heterostructures for $> 10^3$ cycles. Moreover, it is noteworthy that the effective switching speed increases with repeated measurements, implying efficient proton paths are established in the YSZ oxide through repeated cycling.

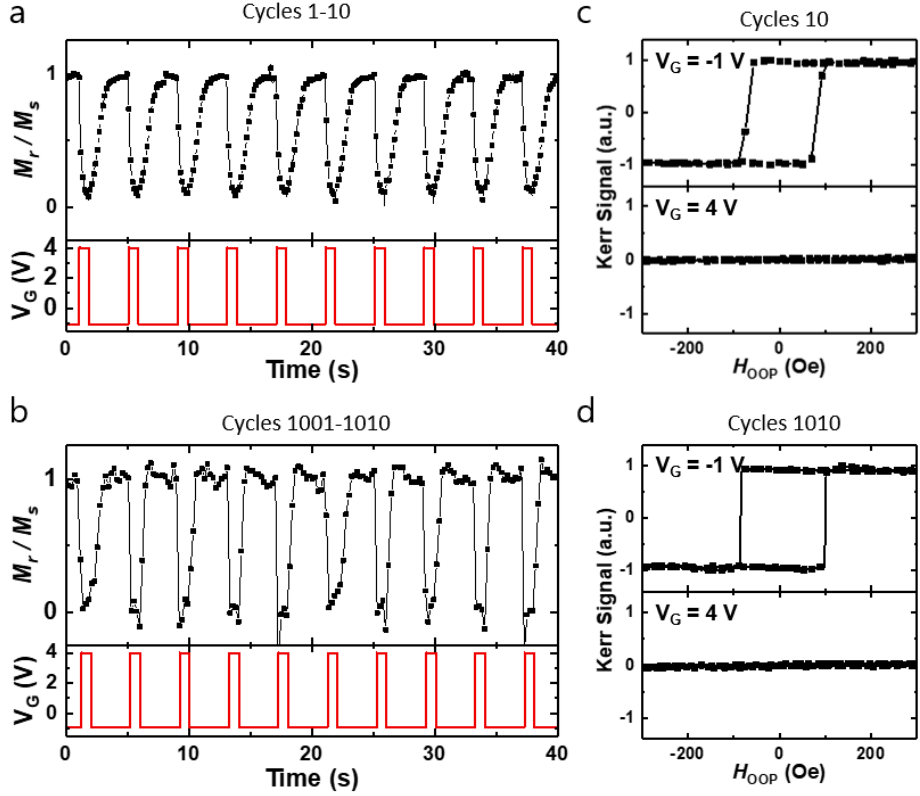

**Figure 4**. **Reliable switching of magnetic anisotropy.** The out-of-plane remnant magnetization ratio ($M_r/M_s$) as a function of voltage cycle for **(a)** cycles 1-10 and **(b)** cycles 1001-1010, respectively. Hysteresis loops under gate voltage measured by Polar MOKE for **(c)** cycle 10 and **(d)** cycle 1010, respectively. Note that $V_G = -1V$ is used here to accelerate the process of in-plane (IP) to out-of-plane (OOP) magnetization transition with limited oxidation of Co.

In conclusion, we have shown that magneto-ionic control of magnetic anisotropy can be highly efficient in ferromagnetic heterostructures using an YSZ gate oxide as a solid-oxide proton electrolyte. Using engineered heterostructure consisting of Pd/Co/Pd/YSZ/Pt, we demonstrate ~millisecond magnetization toggle switching between out-of-plane and in-plane magnetization states. The switching speed can further be engineered by controlling the gate oxide thickness or increasing the voltage amplitude to increase the driving electric field. We also demonstrate robust device endurance with $10^3$ voltage cycles. Together, we believe this work establishes that gate oxide material engineering may be a key pathway for achieving fast and reliable switching in magneto-ionic devices, highlighting the significant potential of this approach towards future spintronic applications.

**Methods**

**Sample preparation.** Ta/(Pt or Pd)/Co/(Pd or not) films were deposited on thermally oxidized Si wafers using a dc-magnetron sputtering at Ar 2 mTorr partial pressure, where the bottom Ta layer serves as a seed layer. Proton conducting oxide (BZCY, GDsC, YSZ) layers were deposited by using pulsed laser deposition (PLD) with a KrF excimer laser ($\lambda$=248 nm) and $O_2$ pressure of ~ 50 mTorr at room temperature, therefore leading to an amorphous microstructure [see Supplementary Fig. S4 for details]. For gate voltage application, 50-μm diameter Au or Pt (3 nm) top electrodes were patterned using shadow masks. The bottom continuous metal layer was exposed at the sample edge and was contacted directly for use as counter-electrode. Cross-sectional transmission electron microscopy (TEM) image of an exemplary Pt/Co/YSZ/Pt heterostructure is shown in Supplementary Fig. S5.

**Polar and longitudinal MOKE measurement.** Polar and longitudinal MOKE measurements were performed using a 660 nm diode laser attenuated to 1 mW. The laser was focused to a spot size of about ~ 3 µm diameter. The gate voltage $V_G$ was applied using a mechanically compliant BeCu microprobe tip. This probe tip was placed near the edge of the top electrode to avoid overlap with the MOKE laser spot position, and the Ta/Pt or Pd back electrode was grounded using another tungsten microprobe tip. The laser spot was focused on the middle of the electrode. For cyclic voltage application experiments, we obtained the hysteresis loop using polar MOKE measurement, where the gate voltage was applied using a function generator. The $M_r/M_s$ value of each point was acquired from individual out of plane hysteresis loops at each measurement recorded at every 2 ms. Note that the time resolution is a limitation of the current measurement setup, and the effective coercivity of the magnetic heterostructure at higher field-sweep frequencies exceeds the maximum available magnetic field amplitude. Moreover, the total number of cycles in voltage switching measurements was limited to ~1,000 cycles, since the mechanical vibration of the system often leads to loss of electrical contact of the BeCu microprobes.

**Cyclic voltammetry measurement.** A Au(10 nm)/YSZ(50 nm)/Pt(3 nm) heterostructure was fabricated for the cyclic voltammetry (CV) measurements. CV measurements were performed using a PARSTAT MC potentiostat (Princeton Applied Research). The top Pt layer and bottom Au layer act as the working electrode and the counter/reference electrode, respectively. The scan rate varied from 10 to 200 mV s$^{-1}$.

**Author Contributions**

S.W. and G.S.D.B. planned and supervised the project. S.J. and D.C. fabricated magneto-ionic devices using oxide materials grown by J.-H.P., H.-I.J. and J.-W.S.. K.-Y.L. and S.J. performed gating experiments with the support from A.J.T., M.H. and J.C.. K.-Y.L., S.W. drafted manuscript and G.S.D.B. revised it with the help from all authors.

**Acknowledgements**

This work was partially supported by the KIST Institutional Program (2E29410, 2E30220), the National Research Council of Science and Technology (NST) grant (CAP-16-01-KIST) by the Korea government (MSIP). This work was also partially supported by National Science Foundation (NSF) through the MIT Materials Research Science and Engineering Center (MRSEC) under award number DMR-1419807, and through NSF award number ECCS-1808828. J.C. acknowledges the support of Yonsei-KIST Convergence Research Institute. S.W. acknowledges management support from Guohan Hu and Daniel Worledge at IBM Research.

**Table of Contents Graphic**

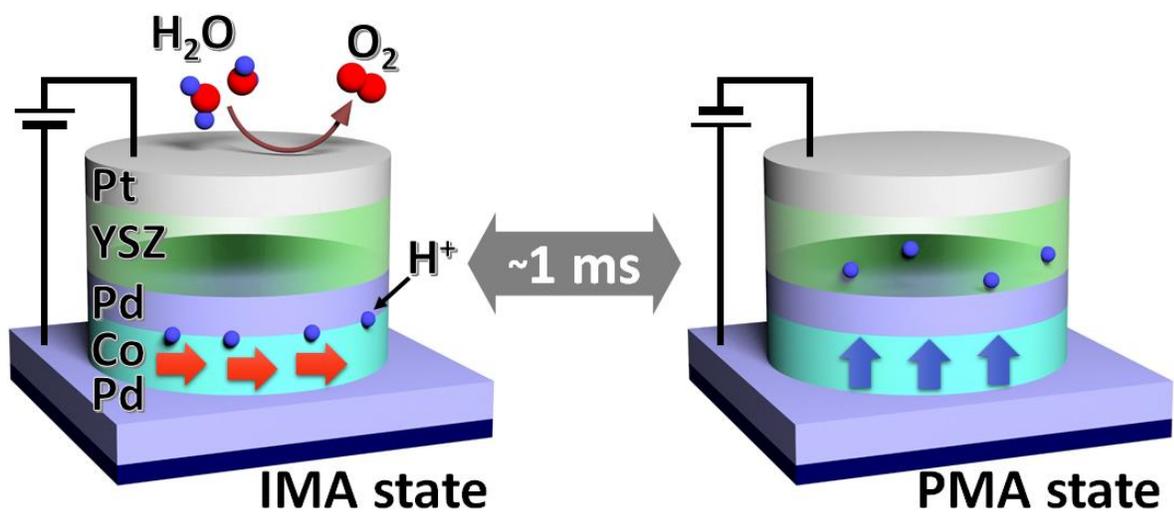